\documentclass{ws-procs9x6}

\begin{document}

\title{The Dissociation of $^8$B in the Coulomb Field \\
and the Validity of the CD Method
\footnote{Work Supported by USDOE grant No. DE-FG02-94ER40870}}

\author{Moshe Gai}

\address{Laboratory for Nuclear Science at Avery Point, 
University of Connecticut, \\ 1084 Shennecossett Rd, Groton, CT 06340-6097.\\
and\\
Department of Physics, WNSL Rm 102, Yale University, \\ PO Box 208124, 
272 Whitney Avenue, New Haven, CT 06520-8124.\\
\     \\
e-mail: moshe.gai@yale.edu, URL: http://www.phys.uconn.edu}

\begin{abstract}
The GSI1, GSI2 (as well as the RIKEN2 and the corrected GSI2) measurements of the Coulomb Dissociation (CD) of $^8B$ are in good agreement with the most recent Direct Capture (DC) $^7Be(p,\gamma)^8B$ reaction measurement performed at Weizmann and in agreement with the Seattle result. Yet it was claimed that the CD and DC results are sufficiently different and need to be reconciled. We show that these statements arise from a misunderstanding (as well as misrepresentation) of CD experiments. We recall a similar strong statement questioning the validity of the CD method due to an invoked large E2 component that was also shown to arise from a misunderstanding of the CD method. In spite of the good agreement between DC and CD data the slope of the astrophysical cross section factor ($S_{17}$) can not be extracted with high accuracy due to a discrepancy between the recent DC data as well as a discrepancy of the three reports of the GSI CD data. The slope is directly related to the d-wave component that dominates at higher energies and must be subtracted from measured data to extrapolate to zero energy. Hence the uncertainty of the measured slope leads to an additional uncertainty of the extrapolated zero energy cross section factor, $S_{17}(0)$. This uncertainty must be alleviated by future experiments to allow a precise determination of $S_{17}(0)$, a goal that so far has not be achieved in spite of strong statement(s) that appeared in the literature.

\end{abstract}


\bodymatter

\section{Introduction}

The Coulomb Dissociation (CD) method was developed in the pioneering 
work of Baur, Bertulani and Rebel \cite{Baur} and has been applied to the case 
of the CD of $^8B$ \cite{Mot94,Kik,Iw99,Sch03} from which the cross section of the 
$^7Be(p,\gamma)^8B$ reaction was extracted. This cross section is essential for 
calculating the $^8B$ solar neutrino flux. The CD data were analyzed with a 
remarkable success using only first order Coulomb interaction that includes only 
E1 contribution. An early attempt (even before the RIKEN data were published) 
to refute this analysis by introducing a 
non-negligible E2 contribution \cite{Lang} was shown \cite{Gai} to arise from a neglect  
of the angular acceptance of the RIKEN1 detector and a misunderstanding of the 
CD method. Indeed the CD of $^8B$ turned out to be a testing ground of the very 
method of CD. Later claims by the MSU group for evidence  
 \cite{MSU} of non-negligible E2 contribution 
in {\bf inclusive measurement} of an asymmetry, were disputed in 
a recent {\bf exclusive measurement} of a similar asymmetry by the GSI2 
collaboration \cite{Sch03}.

In contrast, Esbensen, Bertsch and Snover \cite{PRL} recently claimed 
that higher order terms and an E2 contribution 
are an important correction to the RIKEN2 data \cite{Kik}.
It is claimed that "$S_{17}$ values extracted from 
CD data have a significant steeper slope as a 
function of $E_{rel}$, the relative energy of the proton and the $^7Be$ fragment, 
than the direct result". However they find a substantial correction only to the RIKEN2 
CD data and claim that this correction(s) yield a slope of the RIKEN2 data in better 
agreement with Direct Capture (DC) data. In addition it is stated \cite{PRL} that 
"the zero-energy extrapolated $S_{17}(0)$ values inferred from CD measurements are, 
on the average 10\% lower than the mean of modern direct measurements". The 
statements on significant disagreement between CD and DC data are based on 
the re-analyses of CD data by the Seattle group \cite{Jung03}. In this paper we demonstrate that 
an agreement exists between CD and DC data and the statements of the Seattle group \cite{Jung03} 
are based on misunderstanding (as well as misrepresentation) of CD data.  

In spite of the general agreement between CD 
and DC data, still the the slope of astrophysical cross 
section factor measured between 300 - 1,500 keV can not be extracted with 
high accuracy. This hampers our ability to determine the 
d-wave contribution that dominates the cross section of the $^7Be(p,\gamma)^8B$ 
reaction at higher energies and must be subtracted for extrapolating the s-wave to 
zero energy. Lack of accurate knowledge of the d-wave contribution to  
data (even if measured with high accuracy), precludes 
accurate extrapolation to zero energies. We show that this leads to additional 
uncertainty of the extrapolated $S_{17}(0)$. We doubt the strong statement that 
$S_{17}(0)$ was measured with high accuracy (see for example \cite{Jung03}).

\section{The Slope of $S_{17}$ Above 300 keV}

Early on it was recognized that s-wave capture alone yields an s-factor with a negative 
slope. This is due to the Coulomb distortion of the s-wave at very low distances. 
The observation of a positive slope of $S_{17}$ measured at energies 
above 300 keV was recognized as due to the d-wave contribution. 
It was also recognized that the d-wave contribution is very large at measured energies 
and in fact it dominates around 1.0 MeV. The d-wave contribution must be subtracted 
to allow an accurate extrapolation of the s-wave to zero energy (where the d-wave 
contribution is very small, of the order of 6\%). The (large) contribution of the d-wave at energies 
above 300 keV leads to a linear dependence of $S_{17}$ on energy (with a positive slope). 
An accurate extrapolation of $S_{17}$ must rely on an accurate knowledge of the 
d-wave contribution or the slope at energies above 300 keV.

\begin{figure}
\begin{center}
\includegraphics[width=3.4in]{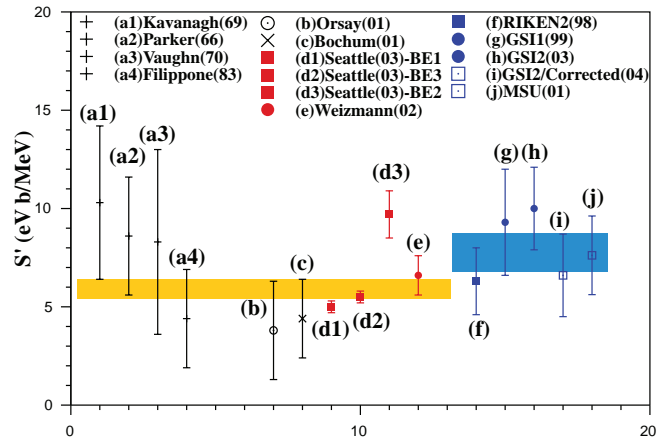}
\end{center}
\caption{\label{SlopeS} The measured slopes (S' = dS/dE) 
of world data measured between 300 and 1500 keV, as discussed in 
the text. The range of "average values" is indicated and discussed in the text.}
\end{figure}

In Fig. 1 we show the slope parameter (S' = dS/dE) extracted from both DC and CD data in the 
energy range of 300 - 1500 keV. We refer the reader to \cite{Gai06} for detailes on data 
used to extract the slope shown in Fig. 1. 
 We conclude from Fig. 1 that the slope parameter can not be extracted
 from DC data \cite{Jung03,Ham01,Str01,Weiz,Fil83,Vaughn,Parker,Kav} 
 with high accuracy as claimed. The 
 DC data are not sufficiently consistent to support this strong statement
  \cite{Jung03}; for example there is not a single data point measured by the 
  Bochum group \cite{Str01} that agrees with that measured by the Seattle 
  group \cite{Jung03}, where we observe that some of the individual data points 
  disagree by as much as five sigma. The disagreement of the three slopes measured 
  by the Seattle group and the disagreement with the Weizmann slope 
  are most disturbing. In the same time 
  the dispersion among slopes measured in CD is also of concern. However, 
  it is clear that the over all agreement between CD and DC data (1.7 sigma) is better than 
  the agreement among specific DC data. We do not support the strong claim of substantial 
  disagreement between slopes measured in DC and CD \cite{Jung03}.
 
 \begin{figure} 
\begin{center}
\includegraphics[width=3.4in]{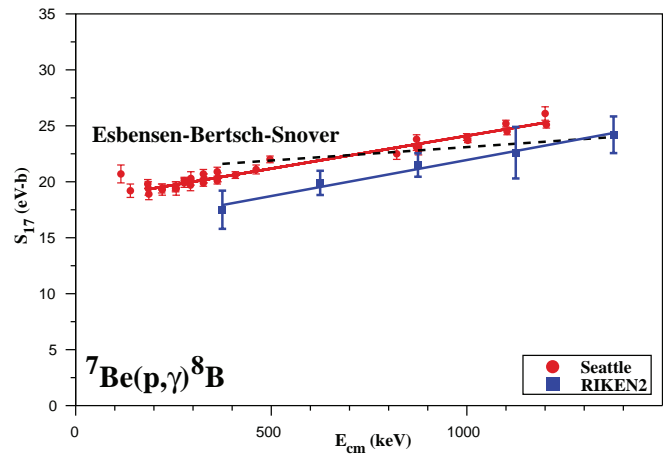} 
\end{center}
\caption{\label{RIKEN2} Extracted $S_{17}$ from the RIKEN2 CD data \cite{Kik} using first 
order electric dipole interaction as shown in \cite{Sch03}, compared to the DC 
capture data published by the Seattle group \cite{Jung03} and the so called reconciled 
slope calculated by EBS \cite{PRL}. The shown RIKEN2 data include 
systematic uncertainties (equal or slightly smaller) as published  \cite{Kik}.}
\end{figure} 

  The lack of evidence for substantial difference between CD and DC results leads to 
  doubt on the very need to reconcile these data \cite{GaiPRL}. Furthermore, in Fig. 2 we 
  show the slope obtained by EBS after their attempt to reconcile the slope of CD with  
  the slope of DC data. Clearly the original slope of the 
  RIKEN2 data obtained using only first order E1 interactions is in 
  considerably better agreement with DC data than the so called reconciled slope.

\section{$S_{17}(0)$ Extracted From CD Data}

In Fig. 20 of the Seattle paper \cite{Jung03} they show extracted $S_{17}(0)$ from CD
using the extrapolation procedure of Descouvemont and Baye \cite{DB}, 
and based on this analysis it is stated \cite{PRL} that "the zero-energy extrapolated 
$S_{17}(0)$ values inferred from CD measurements are, on the average 10\% 
lower than the mean of modern direct measurements". The extracted 
$S_{17}(0)$ shown in Fig. 20 \cite{Jung03} are only from data measured at energies below 
425 keV and the majority of CD data points that were measured above 425 keV 
were excluded in Fig. 20 \cite{Jung03}. 

This arbitrary exclusion of (CD) data above 
425 keV has no physical justification (especially in view of the fact that 
the contribution of the 632 keV resonance is negligible in CD). For example 
as shown by Descouvemont \cite{D} the theoretical error increases to 
approximately 5\% at 500 keV and in fact it is slightly decreased up to 
approximately 1.0 MeV, and there is no theoretical justification for including data
up to 450 keV but excluding data between 500 keV and 1.0 MeV. 

\begin{figure}
\begin{center}
\includegraphics[width=2.9in]{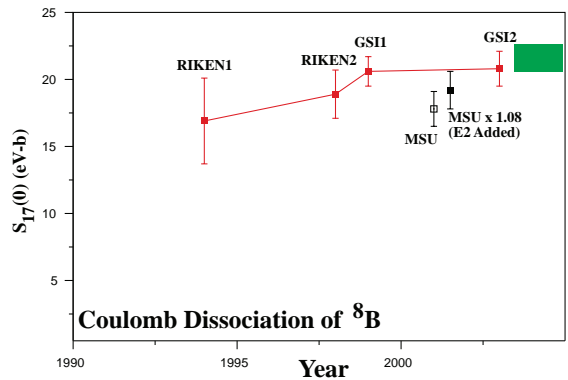}
\end{center}
 \caption{\label{CDDB} Measured $S_{17}(0)$ as originally published by the authors 
who performed the CD experiments. These analyses include all measured data 
points \cite{Mot94,Kik,Iw99,Sch03,MSU} using the 
extrapolation procedure of Descouvemont and Baye \cite{DB}. We also plot 
the MSU data as published as well as with the E2 correction 
($\approx 8\%$) \cite{MSU} added back 
to the quoted $S_{17}(0)$, as  discussed in the text. The range of $S_{17}(0)$ results from the 
measurements of DC by the Seattle \cite{Jung03} and Weizmann groups 
\cite{Weiz} is indicated.}
\end{figure} 

Thus when excluding the CD data above 425 keV, the Seattle group excluded the 
data that were measured with the best accuracy and with smallest systematical 
uncertainty. If in fact one insists on such an analysis of CD data, one must estimate the 
systematic uncertainty due to this selection of data. 
This has not been done in the Seattle re-analyses of CD data \cite{Jung03}.

Instead we rely here on the original analyses of the authors that published the CD
data. In Fig. 3 we show the $S_{17}(0)$ factors extracted by the original authors who 
performed the CD experiments. These results include all measured data points up
to 1.5 MeV, and are analyzed with the same extrapolation procedure of Descouvemont 
and Baye \cite{DB}. 

\begin{figure}
\begin{center}
\includegraphics[width=3.4in]{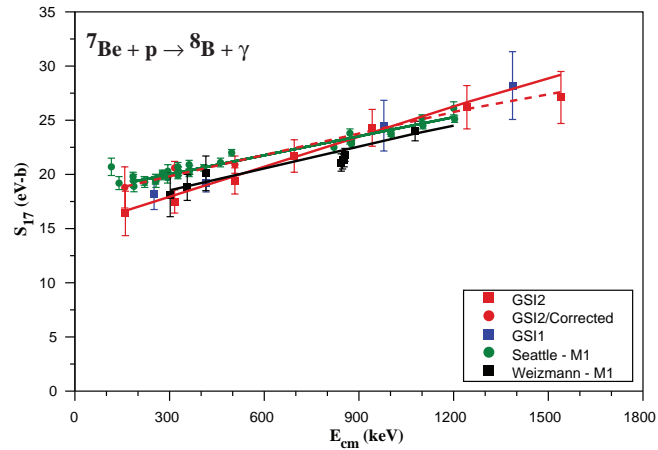}
\end{center}
 \caption{\label{World} A comparison of the most recent DC data with the GSI1 
and GSI2 results.}
\end{figure}

We note that the (four) CD results are consistent within the quoted error bars, 
but they show a systematic 
trend of an increased $S_{17}(0)$ (to approximately 20.7 eV-b), while the error 
bars are reduced. We obtain a  
1/$\sigma$ weighted average of $S_{17}(0)$ = 20.0 $\pm$ 0.7 with $\chi ^2$ = 0.5, 
which is in excellent agreement with the measurement of the Weizmann group 
\cite{Weiz} and in agreement with the measurement of the Seattle group \cite{Jung03}.

\section{Extrapolating $S_{17}(0)$ From World Data}

The current situation with our knowledge of $S_{17}$ and the extrapolated $S_{17}(0)$ 
 is still not satisfactory. The main culprit are major disagreements among DC data. It is clear for example that the systematic disagreements between the Orsay-Bochum \cite{Ham01,Str01} and the Weizmann-Seattle \cite{Jung03,Weiz} results must be resolved before these data are included in a so called "world average". In Fig. 4 we compare the most recent Seattle-Weizmann data (with M1 contribution subtracted) 
with the GSI1 and GSI2 (as well as corrected GSI2) results. While the data appear in agreement
we still observe a systematic disagreement between all measured slopes. The DC data of the Seattle  and the Weizmann groups have different slopes as do the GSI1, GSI2 and corrected GSI2 data. The slope above 300 keV is directly related to the d-wave contribution that dominates at measured laboratory energies, but must be subtracted to extrapolate to solar burning energies. This disagreement does not allow for an accurate (better than 5\% accuracy) extrapolation of $S_{17}(0)$ and must be resolved by future experiments. A reasonable systematic error of +0.0 -3.0 eV-b due to extrapolation seems to be required by current data.

\vfill\eject
\end{document}